## Chapter 22: Magnetic Nanoparticles for Neural Engineering

Gerardo F. Goya[*], Vittoria Raffa

## ABSTRACT

Magnetic nanoparticles (MNPs) are the foundation of several new strategies for neural repair and neurological therapies. The fact that a remote force can act on MNPs at the cytoplasmic space constitutes the essence of many new neurotherapeutic concepts. MNPs with a predesigned physicochemical characteristic can interact with external magnetic fields to apply mechanical forces in definite areas of the cell to modulate cellular behaviour. Magnetic actuation to direct the outgrowth on neurons after nerve injury has already demonstrated the therapeutic potential for neural repair. When these magnetic cores are functionalized with molecules such as nerve growth factors or neuroprotective molecules, multifunctional devices can be developed. This chapter will review some of these new nanotechnology-based solutions for neurological diseases, specifically those based on the use of engineered MNPs used for neuroprotection and neuroregeneration. These include the use of MNPs as magnetic actuators to guide neural cells, modulate intracellular transport and stimulate axonal growth after nerve injury.

## 22.1. Historical Summary and State of the Art

Nerve damage and neurological pathologies are two problems of significant medical and economic impact because of the hurdles of losing nerve functionality as a consequence of nerve injury or degenerative diseases (ND). Nerve regeneration is a complex biological phenomenon.[1] In the peripheral nervous system (PNS), nerves regenerate spontaneously only when injuries are minor. Short gaps can be repaired directly by mobilization of the proximal and distal stumps with end-to-end coaptation and epineural suturing. Long nerve gaps greater than 2 cm require additional material to bridge the defect. The current repair method is the use of autologous nerve grafts (autografts), which provide the regenerating axons with a natural guidance channel populated with functioning Schwann cells surrounded by their basal lamina.[1] Nerve autografting, however, is far from an optimal treatment, and there is suboptimal functional recovery despite technical excellence. These grafts are taken primarily from the sural nerve of the patient. Surveys of the clinical literature show that approximately a half of patients with median and ulnar nerve repairs experience satisfactory motor and sensory recovery.[2] The main reasons for the poor functional recovery rates associated with autografts are unavailability of motor nerves (these grafts are primarily sensory) and mismatch in axonal size.[3] The use of autograft has also the disadvantages associated to the requirement for a second surgical site (donor site morbidity, donor site mismatch and the possibility of painful neuroma formation and scarring)[4]. The use of nerve guidance conduits (NGCs) is the only clinically approved alternative to the autograft for the treatment of large peripheral nerve injuries. They provide a conduit during the nerve regeneration process for the diffusion of growth factors secreted by the injured nerve ends and to limit the injury site infiltration by scar tissue.[5] However, commercially available devices, based on biodegradable polymer or collagen-based hollow tubes, do not match the regenerative levels of autografts, providing good performances only for short defects (< 2 cm) but poor functional recovery for longer nerve gaps.[6] Current knowledge suggests combining the use of NGCs with strategies of molecular or cellular therapies. Molecular therapies deals with the delivery of molecules such as guidance cues (netrins, ephrins, semaphorins and other molecules capable of orientating migrating and growing cells) and factors influencing neuronal growth (e.g. growth factors, neurotransmitters, extracellular matrix proteins)[7]. Cell therapies involve cell transplantation to reduce tissue loss, promote axonal regeneration, facilitate myelination of axons or promote the secretion of factors sustaining the regeneration process. The nanotechnology applied to neural development, repair and protection aspire to implement these approaches for optimal regeneration and recovery of function and to solve those drawbacks arising from the invasiveness of macroscopic implants, including the dependence of the overall performance of such implants to physiological reactions (e.g. fibrosis). Low invasiveness and high selectivity of the growth stimulation are usually conflicting requirements and thus new approaches must be pursued in order to overcome such limitations.

Repairing therapies for those injuries of the central nervous system (CNS) of either the brain or spinal cord are much more challenging. Because the brain coordinates all higher-level functions and communicates with the PNS through the spinal cord, the cellular responses to a mechanical insult and posttrauma situation are numerous, and they are not well understood. Anyhow, once the CNS injury is produced, it initiates a cascade of deleterious events that can affect both cell body and axonal function, resulting in continued dysfunction and prolonged degeneration. For this type of CNS damage, cell therapy is the only therapeutic strategy that proved to work so far. Injured central axons do not spontaneously regenerate. However, the knowledge accumulated during the last two decades challenged the notion that neurons of CNS lack regeneration ability. Although in the 19th century Santiago Ramon y Cajal first suggested the idea that central axons could regenerate but the CNS does not offer a permissive environment, the extended concept of neuroregeneration, including the possibility of neurogenesis and neuroplasticity is a relatively recent one. The notion that neurogenesis is possible led to the idea of implanting viable cells as a therapeutically sound approach in neuroscience. Experimentally, this was demonstrated by transplanting a sciatic nerve explant into optic nerve lesions: the optical nerve regenerated across the graft but growth ceased as soon axons had crossed the graft and reached the interface with the CNS.[8] It is likely that the regenerative potential of central axons is expressed when the CNS glial environment is changed to that of the PNS.[9] It was proposed to bypass the problem by transplanting some specific cell type, which could provide a permissive environment for elongated axon growth, similarly to the Schwann cells of peripheral nerves.[10] Thirty years later this hypothesis was tested for the first time in a 38-year-old male with a complete chronic thoracic spinal cord injury (SCI). This patient received an autologous sural nerve graft to bridge an 8-mm gap and the transplantation of glia olfactory ensheathing cells (OECs) in the proximal and distal nerve stumps; he experienced functional regeneration of supraspinal connections.[11] Some of clinical studies using cell therapy have been or are being conducted for the treatment of chronic SCI,[a] traumatic SCI,[b] amyotrophic lateral

---

[a] www.clinicaltrials.gov. Studies ref. NCT01772810, NCT02688049
[b] www.clinicaltrials.gov. Study ref. NCT02326662

sclerosis,[a] Parkinson's Disease,[b] cervical and thoracic SCI,[c] age-related macular degeneration,[d] etc. These are early stage trials (phase I/II) to assess the safety of the treatment, which is essentially unknown despite a large amount of data available from preclinical experimentation.

Nanotechnology comes into neuroscience to provide additional ways to tackle the above-mentioned problems. Since nanoparticles (NPs), and more generally nanostructures, can be made small enough to interact with subcellular structures, the possibilities of intracellular targeting and actuation on damaged neural cells are countless. Inorganic NPs can be engineered as drug carriers alone for releasing neuroregenerative drugs, as reported using hollow silica NPs with porous walls to control the drug release kinetics.[12] Also, different physical properties of the NP's core or coating can be used to trigger the release, providing spatial and temporal control of the dose. The possibility of surface functionalization of NPs adds potential increase of specificity and/or hydrophobicity solutions for already existing therapeutic drugs. Among these strategies, the use of noncontact forces such as magnetic fields provides alternatives for remote NP actuation and activation. This chapter will focus on the new solutions nanotechnology can provide for neurological diseases, through engineered MNPs applied to neuroprotection and neuroregeneration. Also, the application of MNPs as magnetic actuators to position or guide neural cells by an external magnetic field will be described and discussed. In the first part we have included a description of the magnetism related to MNPs, as well as the theoretical framework for magnetic field interactions with biological systems. In the second part of the chapter, we offered an outline of the different strategies based on the use of MNPs and magnetic fields, applied to (a) neuroprotection in neurodegenerative diseases and (b) nerve regeneration following injury. We also describe and discuss those relevant MNP-based strategies successfully employed to remotely guide neuronal growth under the action of magnetic fields.

---

## 22.2. Magnetism of Single-Domain Nanoparticles

The possibility of remote actuation on a nanoscale object has been understood since long ago as a way to manipulate biological systems.[13] From the physical point of view, the action-at-a-distance is a consequence of the interaction among any magnetic dipole (the basic entity in magnetostatics) having magnetic moment **m** and the magnetic flux density **B**, also known as magnetic induction.[14] In the general case, the force on a magnetic moment **m** exerted under a magnetic induction **B** is given by the expression

$$\mathbf{F} = \nabla(\mathbf{m} \cdot \mathbf{B}), \qquad (22.1)$$

where the spatial derivative implies that a nonuniform field is required to apply forces. In addition, the **B** field exerts a torque $\mathbf{N} = \mathbf{m} \times \mathbf{B}$ on the magnetic moment **m** that will align the dipole parallel to **B**. Therefore, for those applications that require maximizing the magnetic forces between the external field **B** and the magnetic moment **μ** of the MNPs, the usual strategies rely on (a) the design of optimized magnetic field profiles, and (b) the synthesis of MNPs with large magnetic moments. The former choice is rather old and there is an extensive bibliography on numerical methods and magnetic field configurations.[15–17] For a comprehensive review of nanomagnetism and magnetic properties of MNPs the reader is referred to the comprehensive work of D. Ortega (Chapter 1: Structure and Magnetism in Magnetic Nanoparticles in the book *Magnetic Nanoparticle: From Fabrication to Clinical Applications*).[15]

The choice of the material for the magnetic core of MNPs is related to the physical and magnetic properties of the corresponding bulk phase. However, below a given critical particle diameter $d < d_{crit}$ (with $30 \leq d_{crit} \leq 100$ *nm*, depending on the material's nature) the magnetic structure of the particle's core is different than the bulk material in the sense that domain walls collapse into a single magnetic domain. A deeper analysis of the concepts of magnetic domains, magnetic order in small particles and superparamagnetism is beyond the scope of this chapter, and the reader is referred to Ref. [18] and the classic book by B.D. Cullity[16] (Chapter 8). The value of $d_{crit}$ is determined by the magnetic anisotropy ($K$) and the exchange stiffness coefficient ($A$) of the bulk material, and $d < d_{crit}$ defines a size regime below which the magnetic cores are magnetically ordered in a single direction. Therefore, this spin alignment results in a net magnetic moment of several hundreds of Bohr magnetons (Bohr magneton is the elementary unit of magnetic moment, defined in SI units in terms of the electron charge $e$, and mass me, and the reduced Planck constant $\hbar$, by $\mu_B = e\hbar / 2m_e$ ). The magnetostatic energy of a single-domain MNP (i.e. the magnetic energy in the presence of an externally applied magnetic field) is proportional to its volume V, and this energy competes with the thermal energy to keep the magnetic moment spatially fixed.[16] Around room temperature (i.e. within the 25–45°C range), where most biomedical uses occur, the thermal energy can be of the same order than the magnetostatic energy for small applied fields. Therefore, the thermally induced magnetic relaxation impairs the magnetic alignment of **m** and **B** diminishing the magnetization at low fields. For MNPs with average size < 30 nm thermal relaxation is predominant and thus affects the efficacy of those biomedical applications that require full magnetic saturation at room temperature. For these applications, the design of MNPs must consider average particle size and/or magnetic anisotropy large enough to prevent thermal relaxation.

### 22.2.1.  Magnetic Field–Magnetic Nanoparticle Interactions

The strategy of using MNPs to actuate cells mechanically can be traced back to the year 1920, when W. Seifriz[17] proposed the use of 'minute particles of magnetic material' to measure the elasticity of the cell cytoplasm. Since then, a large amount of theoretical and experimental work on magnetically loaded cells has been reported.[18] The physical concept behind this approach is based on the interaction between the magnetic (dipole) moment **m** of MNPs and a spatially inhomogeneous magnetic field **B,**[19] as described by Equation 22.1. For a spherical MNP composed of magnetite, $Fe_3O_4$, with diameter $d$ = 50 nm, a magnetic moment of $m \approx 7 \times 10^{-17} Am^2$ can be estimated.[20] Assuming a commercially available NdFeB magnet (e.g. type N50) of cubic shape with dimensions 1×1×1 $cm^3$, a single MNP located at a distance of 5 mm from the surface will experience an average force $F \approx 2.1 \times 10^{-15} N$. This force is larger than the gravitational force $(\sim 10^{-18} N)$.[21] In addition, biomedical applications imply that the MNPs are immersed in a fluid and therefore the Stokes law predicts that any particle moving with velocity $\vec{v}$ will experience a drag force $\vec{F}_D$ given by $\vec{F}_D = 6\pi\eta r\vec{v}$, where $\eta$ is the viscosity of the medium and $r$ is the radius of the MNP. This force is size-dependent, but for the applications in quasi-stationary conditions such as those existing in a cell culture, the velocity factor makes this force small enough to discard it.[22] On the other hand, diffusional forces due to Brownian motion are also size dependent and cannot be neglected in colloidal systems at room temperature. A complete analysis of the influence of Brownian forces of single-domain MNPs under external magnetic field requires the use of stochastic approaches, such as the stochastic Eulerian–Lagrangian method, which is beyond the scope of the present description. For a more complete presentation the reader is referred to the recent works in Refs. [26] and [27]. Summarizing, to produce a measurable pulling-magnetic force, the MNPs have to be designed to maximize their magnetic moment $\mathbf{m}$ under the $\mathbf{B}$ values applied, and the magnetic field profile must be planned to produce enough field gradients for the experimental conditions required.[23]

### 22.2.2.    Physical Features of Magnetic Nanoparticles

Two iron oxides, namely magnetite ($Fe_3O_4$) and maghemite ($\gamma$-$Fe_2O_3$), are by far the most used materials as a constituent of MNPs in the biomedical field.[24] These oxides crystallize in the cubic spinel structure, where two cationic sites with different geometries define the two magnetic sublattices, labelled as A and B sites. The cations at A and B sublattices have different atomic magnetic moments and result in a noncancelling total magnetic moment, and this type of magnetic order is known as ferrimagnetism. The macroscopic behaviour is similar to a ferromagnetic one, with remanence (i.e. a net magnetization at zero applied field), hysteresis (i.e. magnetization dependence on the magnetic history) and magnetic ordering (Néel) temperature. As mentioned before, in MNPs with size below the critical domain size the formation of domain walls is energetically unfavourable and the magnetic state has a single-domain configuration. If the MNP's volume is small enough, the thermal fluctuations at room temperature makes the magnetic moment to relax within timescales shorter than the measuring time, yielding a null-average of the magnetic moment.[25] This state is known as superparamagnetism.

The magnetic relaxation of MNPs in magnetic colloids is therefore governed by the dynamics of the magnetization vector and have been modelled by Usov et al.[26] using the stochastic Landau–Lifshitz model.

As mentioned before, the specific properties of a given magnetic material must be considered when using MNPs as pulling agents. Specifically, those aspects governing the interaction expressed by Equation 22.1 between the magnetic field $\mathbf{B}$ and the magnetic moment $\mathbf{m}$ of the material. Since the material properties of the MNPs enter Equation 22.1 through the magnetic moment $\mathbf{m}$, it is expected that the resulting magnetic forces should be more or less independent on the physicochemical environment of the MNPs for a given application. Instead, the most relevant parameters are the saturation magnetic moment $M_S$ of the particles and their magnetic anisotropy. Together with the average volume $V$, these parameters will define the magnetic response of the material under the magnetic field intensity $H$ applied. If the MNPs have a single-domain configuration, the magnetic moment measured under a given $H$ value is given by

$$M\left(T,H\right)=M_S\ \mathsf{L}\left(M_S V H / k_B T\right),\qquad(22.2)$$

where $\mathsf{L}\left(x\right)$ is the Langevin function, and $k_B T$ is the thermal energy factor at temperature $T$. This expression, together with Equation 22.1, show that if the MNPs are small enough the thermal fluctuations will render the magnetization small therefore decreasing the magnetic force. On the other hand, for multidomain MNPs the magnetization is governed mainly by domain wall motion and for large $H$ values also by magnetic moment rotation within domains. Therefore, the preferred materials that provide magnetic saturation at low fields (and therefore large magnetic forces) would be magnetically soft materials (i.e. low magnetic anisotropy) without crystalline defects or vacancies, to avoid domain wall pinning.

### 22.2.3. *Instrumentation: Simulation and Application of Magnetic Forces*

Experimentally, the **B** profiles required for *in vitro* experiments can be produced by a suitable conuration of permanent magnets (e.g. FeSm$_5$ or NdFeB magnets) as well as different types of electromagnets. In most biomedical applications, processing conditions require working within fluidic phases and in small volumes. These prerequisites combine well with the use of microfluidics as a complementary technique to handle highly stable microflows and to fit the small volumes of liquids like culture media for *in vitro* experiments, where the total volumes can be as small as 10$^{-9}$ L. The downscaling of magnetic separators allow to integrate them into more complex systems like detection devices for diagnostics and clinical assays, environmental monitoring, food-contaminant analysis, etc.[27],[28],[29] Also, small sample spaces allow larger field gradients to be applied without the need of large magnetic fields.

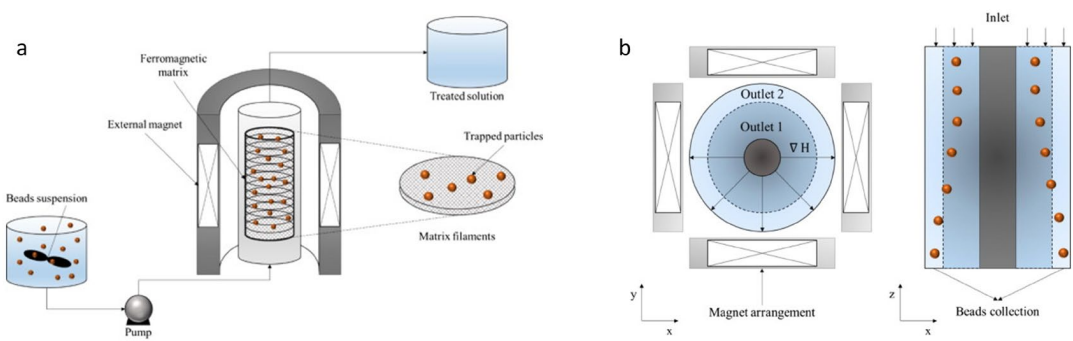

**Figure 22.1:** Schematic drawing of two approaches for magnetic separation under continuous flow sorting (a) a high gradient field separator based on superconducting magnets and (b) a magnetic quadrupole configuration, frequently used in small-volume applications. (Reprinted from *Sep. Purif. Technol.*, 172, J. Gomez-Pastora, X. Z. Xue, I. H. Karampelas, E. Bringas, E. P. Furlani, and I. Ortiz, 16, Copyright 2017, with permission from Elsevier.)

On the other hand, for larger working spaces the use of high-power electromagnets seems to be the only workable choice. One of the possible arrangements to combine continuous sorting and large working volumes is schematized in Figure 22.1a, where a superconductor coil of cylindrical symmetry surrounds a ferromagnetic matrix immersed in the flowing medium. Here the external field provides the large intensity of **B** inside the tube, whereas the ferromagnetic network provides the local inhomogeneity (i.e. field gradient) to retain the magnetic particles. The scalability of magnetic separation by magnetic forces is technically simple, although the amounts of energy required at industrial scales make it expensive. In any case, the use of high-gradient magnetic fields is being used successfully for treatment of industrial wastewaters and removal of heavy metals.[30]

At small working volumes (i.e. *in vitro* or small *in vivo* applications) the adequate choice of the **B** source will depend on the specific details of the experimental setup, but in most cases commercially available permanent magnets can produce suitable magnetic field gradients of several thousand *T/m*. A simple quadrupole configuration used for magnetic separation is produced by four permanent magnets placed on the external side of a supporting tube through which the colloid is pumped (see Figure 22.1b).[31] This configuration produces four regions with maximum field along the circular perimeter of the tube, whereas **B** = 0 at the centre.

A similar approach than the one used for magnetic separation, i.e. the use of the magnetic forces between external dc fields and the MNPs magnetic moment discussed above, is the basis for magnetic applicators designed for magnetic targeting. However, for MNPs to be concentrated at any internal body space there are additional difficulties. First, any realistic *in vivo* situation should consider not only the dynamic nature of the circulating blood but also the nonlinear character of the systemic paths that will carry the MNPs. In addition, the inherent pulling nature of the magnetic forces makes difficult to direct a net magnetic force towards an inner body volume using an external array of magnetic field sources. A potential solution to this problem was proposed through active targeting and accumulation of magnetic actuators to neural cells. This strategy has been successfully applied to control the mammalian nervous system in mice.[32]

## 22.3.  Magnetic Actuation on Neural Cells

### 22.3.1.  *Effects of DC Magnetic Fields on Neural Cells*

A substantial portion of the early research on biomagnetism was devoted to elucidating the influence of static and alternate magnetic fields at cellular and tissue levels.[33] Such investigations have disclosed many biochemical pathways that are influenced by a magnetic field. Only a small number of those investigations were related to physiological mechanisms in vertebrates under the influence of static magnetic fields, describing how reactions to magnetic stimuli were effected through the CNS.[34] The physical mechanism by which an exogenous magnetic field affects the biological pathways in eukaryotic cells is still under discussion, although there is long-standing experimental evidence that demonstrates the measurable effects on cell proliferation, migration and adhesion.[35] The existence of the earth's magnetic field ($H$ = 39.8 A/m or 500 mG) provides examples of biological interactions that are well documented in bees, pigeons, bacteria and fish. The phenomena involving the capacity of a living organism to perceive or detect such weak magnetic field is known as magnetoreception.[36]

Also, the effects of intense static magnetic fields (i.e. up to several MA/m, or kGauss) have been studied in several different animal species, with different results, a relation between long-term application of strong static fields and biological pathways has been suggested. For example, experiments in young mice subjected to strong DC magnetic fields (i.e. $H$ = 334 kA/m or 4200 G) have demonstrated measurable effects including growth retardation, changes in the population of bone marrow-derived monocytes, and increased rates of appearance of spontaneous cancer.[37]·[38]

As mentioned above, there is abundant experimental evidence that the application of static (or very low-frequency) magnetic fields on eukaryotic cells affects many biochemical pathways significantly, including cell proliferation, adhesion[35] and expression of heat-shock protein.[39] In the case of neural cells, the influence of magnetic fields could be expected on those mechanisms involving the exchange of ions through the cell membrane. Theoretical explanations[40] for these effects were proposed through perturbation effects of the magnetic field on moving charges. Since these neural communication mechanisms involve electrical signaling through ion channels at the cell membrane, it seems reasonable to expect that magnetic fields can influence the dynamics of cross-membrane ion pumping, impacting on cell differentiation and cell growth. However, there is experimental evidence excluding measurable effects on Na+ and K+ transmembrane currents down to one part in 1000.[41] On the other hand, it has been suggested that changes in nerve activity when exposed to strong DC magnetic fields (e.g. >100 kA/m) could be related to the diamagnetic anisotropy of some molecular components of the cell membrane. Under high magnetic fields, it is expected that the anisotropy axis of the membrane molecules will align along the field direction, and this realignment would suffice to modify the ion channel activity.

It is interesting to note that the two mechanisms differ on their physical basis: the action of $\vec{B}$ on moving charges $\pm q$ is the Lorentz force $\vec{F} = \pm q\left(\vec{v} \times \vec{B}\right)$ applied to those charges with velocity $\vec{v}$, whereas the diamagnetic alignment of membrane molecules is the response of the closed-shell orbital atomic moments to the applied magnetic field. These differences make it in principle possible to design experiments to identify which mechanism will contribute under specific conditions. Both effects could be significant under strong fields, but the different B-thresholds at which these mechanisms start to operate and to what extent they are independent remain to be elucidated. In any case, the experimental evidence supporting the influence of static magnetic fields on neural cells is already quite solid, and explains why most reports on clinical effects of magnetic fields refer to the nervous system.[42]

Due to the complex interaction between electric and magnetic phenomena, the disentanglement of each source when a given (electrical or magnetic) experiment is performed is always challenging. The classification of 'pure' magnetic or electrical stimulation can be useful sometimes but the electromagnetic theory makes this distinction unfitting in the sense that a 'pure' static field $B$ can modify the distribution of electrical charges existing in any material. Regarding biological materials (e.g. membranes, tissues, body fluids) it has been shown that the most influential physical parameters to be considered to affect cell functions are the electric and magnetic field amplitudes ($E_0$ and $H_0$, respectively), the intensity of induced currents, the induced voltage and the frequency.[43] In any case, general considerations indicate that the time scale of the electromagnetic stimulus must be of the same order of magnitude than the physical mechanism involved because otherwise the time average of the shorter magnetic pulses on the much larger time scales of biochemical dynamics in a cell membrane would produce a null effect. out by simple time averaging any effect. For this reason both DC and extremely low-frequency magnetic fields are the usually chosen regimes to influence the response of biological systems.

Some works published in the 1980s about a 'cyclotron resonant effect' attempted to link weak electromagnetic fields to an enhanced $Ca^{2+}$ transport through cell membrane due to resonant mechanisms.[44] However, attempts to replicate this effect were unsuccessful.[45] Moreover, theoretical considerations about the influence of viscosity and molecular collisions in fluid biological media seem to preclude any possibility of resonance associated with ion trajectories in such magnetic fields.[46]

Iron is a relatively abundant element in most living organisms. Therefore, it is not surprising that biomineralization of iron, i.e. the biochemical processes through which an organism synthesizes hard minerals have made magnetite ($Fe_3O_4$) ubiquitous across both kingdoms of prokaryotes and eukaryotes including bacteria, protozoa and mammals. The occurrence of $Fe_3O_4$ crystals in the human brain resulting from iron biomineralization was first reported by Kirschvink,[47] who showed the presence of 20–50 nm crystals both isolated and forming linear structures similar to those typical of magnetotactic bacteria. The presence of nanostructured magnetite in the brain has been related to NDs in which disruption of normal iron homeostasis occurs.[48] The excess of iron and senile plaques found in brain tissue seem to support this idea.[49] It is interesting to mention that a recent study has suggested airborne pollution as an exogenous source of the $Fe_3O_4$ NPs found in brain tissue,[50] which poses the question of whether the major sources of MNPs in the brain have an internal or external origin. In any case, the idea that these magnetite MNPs within the brain could have relation with some of the biological effects related to AC magnetic fields in humans[51] merits further investigation.

### 22.3.2. *Magnetic forces can actuate on cells*

Although magnetic fields do have an influence on neural tissue, it is evident from the previous discussion that the nature of the interaction makes difficult to envisage their uses for remote tethering or actuation. Magnetic actuation is the action of influencing the behaviour of a cell by magnetic forces, generated from MNPs previously uploaded/attached to the cell.

To have the capacity of influencing axonal growth, magnetic forces must produce an effect larger than the drag forces within the cell, even at the nanometric scale. Magnetic forces originate in the interaction between the magnetic moment of MNPs and the magnetic field, as already discussed in Section 22.2.1, together with drag forces. For cell actuation a way to overcome the effects of drag forces is through the design of the MNPs. Furthermore, novel therapies that use exogenous cells (cell therapy) to gain lost functionalities in target tissues or organs have been proposed,[56–59] which provide a fascinating tool for concurrent uses for MNPs. For example, stem cell-based treatments have been established as a clinical standard of care for some conditions, such as hematopoietic stem cell transplants for leukaemia and epithelial stem cell-based treatments.[52] Although the scope of potential cell-based therapies has expanded in recent years due to advances in basic research, attempts to develop a cell-based intervention into an accepted standard of medical practice are particularly difficult processes for different reasons. One of the unresolved issues relating to the clinical use of transplanted cells concerns the localization of these cells to the diseased site,[61–63] since only a small percentage of the implanted/injected cells *in vivo* reach the desired location.[53]

There is enough evidence that for neural or neural precursor cells MNPs can be incorporated into the cytoplasm in large amounts. For example, the iron uptake in the oligodendroglial cell line OLN-93 has been reported[54] to increase the contents of intracellular iron up to ≈200 times the basal concentration in a concentration-dependent way. A comparative study on internalization in primary and immortalized cells showed that immortalized PC12 cells have a more intense activity than primary cells regarding MNP uptake.[55] The same study revealed that in a mixed (neuronal and glial) primary cell culture the predominant uptake of MNPs was done by microglia, whereas the number of astroglia and oligodendroglia incorporating MNPs was lower. Moreover, comparison against organotypic cocultures of spinal cord and peripheral nerve grafts yielded MNP-uptake levels similar to those of the primary cell cultures.[55]

The way by which a MNP is delivered to the cytoplasmic space can be very different depending on the type of cells or MNPs involved. Little work has been reported on the mechanisms of MNP uptake by neural cells and, more generally, about the interactions between MNPs and neural cell lines. Tay et al. reported a meticulous study on the interactions of MNPs with primary cortical neural networks in different developmental stages.[56] These authors found that chitosan-coated MNPs were internalized whereas starch-coated MNPs were not, the latter being attached to the cell membrane. By inhibiting selectively different uptake mechanisms, they concluded that the mechanisms by which chitosan-coated MNPs were incorporated was micropinocytosis and clathrin-mediated endocytosis.

The latest evolution of nanoscience into the neuroscience field has provided incipient solutions for the remote guidance of functional cells related to the above-mentioned cell therapies.[57, 58] The ability to introduce MNPs into cells and magnetize them was the first step towards remote manipulation by magnetic fields to carry healing cells to the desired site, enabling the cells to colonize and differentiate into any desired cell type.[59] Also, different approaches based on magnetic forces to destroy target (cancer) cells have been reported. For example, Kim et al.[60] have succeeded in provoking cell damage using magnetic microdisks that could be forced to rotate by an external magnetic field of very low frequency (i.e. a few hertz) due to their vortex structure. The mechanical rotation was reported to compromise the integrity of the cell membrane, triggering an apoptotic mechanism. More recently, the same concept has been successfully applied *in vivo* to reduce an intracranial glioma tumor with no observed side effects.[61]

However, the rationale for remote guiding of axonal growth includes not only the successful uptake of the MNPs by the target cells. Given the large number of cells that participate in the repair after nerve injury, there is the question of whether some specific cell types could be more efficient in internalizing the MNPs injected than the target neurons. Most of the previous reports about the effects of NPs on neural cells (e.g. cell uptake, toxicity, etc.) have been conducted on immortalized cell lines (see for example Refs. [73,74]) and only a few studies have been performed to investigate the effects of MNPs on primary cells of the nervous systems.[62]·[63]

The design of any magnetically guided axon regeneration therapy must consider how the external magnetic forces will act on an MNPs-loaded cell. The basis of the remotely guided neural regeneration involves (a) physical mechanisms to direct axonal regrowth along selected directions, and (b) biochemical mechanisms to stimulate axonal elongation across the nerve lesion site.[64] Also, the molecular guidance of axonal growth based on high-affinity molecules (such as growth factors and extracellular matrix proteins) can orientate growing cells,[65] although no therapeutic outcome has yet been reported.

Regarding physical guidance, autologous and heterologous tissue grafts or bioderived materials as scaffolds have been partially successful in providing growth conduits to guide the nerve during regeneration.[66]·[67] On the other hand, the uses of contactless magnetic forces have been much less studied. Some studies have been reported to be effective in both axon orientation and growth,[68]·[68] although these results are up to now limited to *in vitro* experiments. For *in vitro* situations, there are specific adhesion forces chemically sticking the cells to a substrate,[69] and therefore interaction of MNPs with H must be strong enough to overcome the adhesive forces, which have been reported to be in the 1–200 pN range depending on the measuring conditions.[70]

## 22.4. Nerve Repair

### 22.4.1. *Magnetic guidance*

The concept of magnetically assisted nerve repair is based two complementary actions that can be performed when MNPs are used. The first action is related to the interactions with a remote magnetic field, which generates a pulling force from the MNPs to the growing axons. The second, i.e. the possibility of having neurotrophic factors on the MNP's surface, would make possible to stimulate growth rates during the application. This synergistic approach for nerve repair using MNPs is schematically illustrated in Figure 22.2. It is evident that the complexity of the actual biological process, which includes a plethora of different cell types acting on the injured nerve, makes it necessary to verify some hypotheses regarding the effects and fate of MNPs once they are injected into the injured tissue. For example, Schwann cells are recognized as helpful agents promoting axonal regeneration in the PNS while astrocytes and oligodendrocytes in the CNS are not. Therefore, the successful delivery of MNPs to the target axon will depend on the relative affinity of these cell types for the MNPs.

Although mechanisms involved in axonal growth are not completely understood, there is increasing evidence that mechanical force generation is a crucial process for both axonal guidance and lengthening.[71] The existing literature suggests that neurons and their axons possess fine sensors to sense and transduce mechanical force in axon initiation/elongation/guidance.

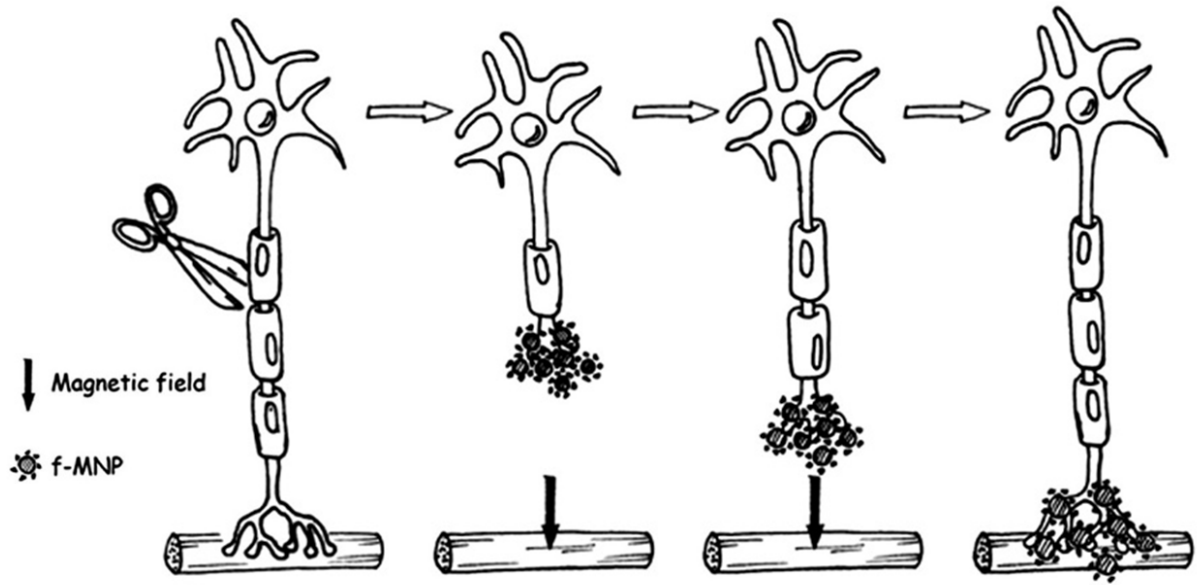

**Figure 22.2:** Schematic illustration of magnetically assisted nerve repair process. The injured nerve is targeted by the magnetic NPs, and then a remote magnetic field guides axonal growth along the field lines. The particles, in turn, could be surface-functionalized with growth-stimulating molecules to accelerate the healing. (Reprinted from *Nanomed-Nanotechnol,* 10, C. Riggio, et al., 1549, Copyright 2014, with permission from Elsevier.)

The involvement of mechanical tension in the morphogenesis of the nervous system was clear in the late 1970s when pioneering experimental work revealed that neuronal processes *in vitro* are under tension.[67] Later, different teams demonstrated that the external application of mechanical tension alone is sufficient to initiate *de novo* axonal sprouting.[84–86] There is a consensus that neurite elongation is a linear function of the applied force and its rate has been found to be similar to both PNS and CNS (about 0.1–1 µm h$^{-1}$ per pN of applied force).[85–88] MNPs, which develop a strong magnetic force when an external magnetic field is applied, could be used to induce an extremely rapid regeneration of the injured axons, purely directed by mechanical forces on MNP-labelled axon tracts. Fass and colleagues used magnetic beads to precisely develop forces in the piconewton range, finding cells able to sustain mechanical-driven elongation with applied tensions between 15 and 100 pN.[72] In addition to the evidence that mechanical tension can induce elongation of neurite or process initiation, recently its influence on axonal guidance has also been investigated. It was demonstrated in a neuron-like cell line that MNPs can be used to gain control of directional movements of neurites. Specifically, by using magnetic nanobeads, it was found that the application of 0,5 pN force on cell neurites was enough to preferentially align them along the direction imposed by the mechanical force.[7] Moreover, using a model based on the effects of the applied forces acting on the receptor–ligand bond, dynamic process of bond loading, breaking and formation during cytoskeletal movements, the authors could reproduce the experimental data successfully.[64] A basic setup for this experimental approach using four parallel NdFeB magnets is depicted in Figure 22.3, where the micrograph (inset) shows the preferential growth along the field lines (yellow arrow), quantified through the angle θ between H and the direction of the main dendrites.

The ability to generate mechanical tensions on neurites, to promote elongation, and to guide directional movement could make MNPs a powerful strategy to address the dream of axonal re-innervation from the CNS to the desired target, e.g. the neuromuscular junction. The MNP-mediated mechanical force has also been used to manipulate neuronal compartments. MNPs have been used to bind filopodia cell membrane of retinal ganglion cell growth cone and to elicit axonal growth and guidance by exerting mechanical tension with an externally applied magnetic field.[73] MNPs functionalized with TrkB agonist antibodies have been used to target the particles to signaling endosomes, to manipulate them by focal magnetic fields, and to alter their localization in the growth cone, thus deregulating growth cones motility and neurite growth.[74] The synaptosomes of brain nerve terminal labelled with MNPs were spatially manipulated with external magnetic field without affecting the key characteristics of glutamatergic neurotransmission.[75] Recently, manipulation via MNP has also been performed at molecular level, by influencing protein segregation during axonal development, *in vitro* and *in vivo*, to dictate axon formation.[76] In general, MNPs offers the distinct advantage of being easily functionalized with ligands for high affinity binding to specific neuronal cell types, compartments or proteins,[77] which makes particularly effective present and future strategies of neuronal manipulation via MNP-induced mechanical forces. MNP have been used also to manipulate the extracellular environment, which plays a key role in the process of nerve regeneration. Recently, magnetic particles have been used to orientate collagen fibres under an external magnetic field, opening the possibility to develop oriented scaffolds to strongly promote the process of functional reinnervation.[78]

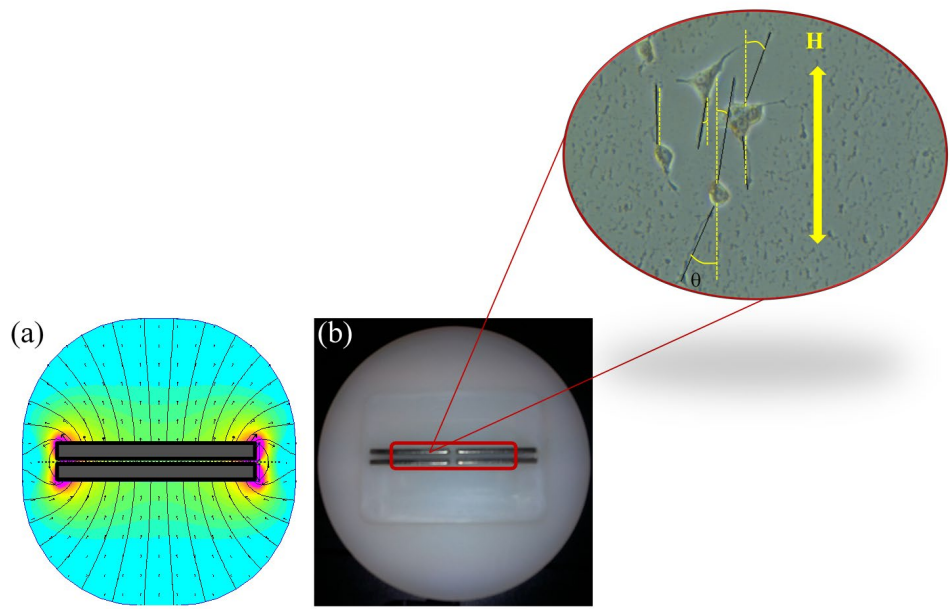

**Figure 22.3: (a)** Representation of the magnetic field applied to the neural PC12 cell cultures. The magnetic field was homogeneous in the Y and X direction (0.19–0.20 T). The maximum magnetic field gradient was 0.019 T/m. (b) Image of the support where the T-25 flasks were incorporated and an example of the images obtained by an optical microscope in the area where the cells are analyzed. The image shows the analysis of the neurite direction; each neurite is manually traced and then the angle formed between the neurite and the direction of the magnetic field (θ) is recorded.

### 22.4.2 Neuroprotection

Functionalization of MNPs with neurotrophic factors to promote neuron survival/growth can also be achieved.[79] Although the free growth factors have a very short half-life (e.g. few minutes)[80] *in vitro* studies have proved that the conjugation to iron oxide NPs can prolong the biological activity of NGF, glial cell-derived neurotrophic factor (GDNF) and basic fibroblast growth factor (FGF-2).[80] [81] An additional advantage of MNPs is that they can be remotely guided by magnetic forces. They have also been used as magnetically guided nanocarriers for spatially controlled drug delivery, e.g. for local release of anaesthetics for local nerve block[82] or for targeting neurotrophic factors to the blood–brain barrier (BBB).[83] The idea of improving neuroprotection using drug-loaded nanocarriers through the BBB is many years old.[84] The capability of a nanometre-sized device with a therapeutic payload to cross the BBB is appealing since about 95% of the therapeutic drugs for treating CNS diseases fail to do so in the brain.[85] This is mainly related to the impenetrability of the BBB for such molecules. Several strategies to overcome this problems using MNPs have been reported, based on the functionalization of the particles with peptides, proteins and similar small molecules.[83] However, the actual neurotoxicity levels of MNPs *in vivo* are not yet completely known. It has been reported that MNPs entering into the body fluid system can result in adverse effects on the CNS.[86] Also, systemic administration of MNPs has been reported to induce breakdown of the BBB, an effect not only exclusive of magnetic particles but NPs in general.[87] Different interactions between MNPs and CNS in physiological vs. pathological conditions cannot be also excluded. A recent work showed that NPs can target myeloid cells in epileptogenic brain tissue, suggesting their use for detecting immune system involvement in epilepsy or for localization of epileptic foci.[88] A related, more subtle question of whether MNPs influences the physiological brain responses under pathological conditions has been addressed only rarely in the literature, but is certainly a subject that merits investigation.

*22.4.3 Magnetofection*

The concept of transfection can be defined as the procedure by which any type of genetic material from a foreign source is introduced into a different mammalian cell. When dealing with DNA, this process enables the expression of proteins from the original source by the host cell's machinery. The transfer of the genetic material can be done by different methods, in many cases using coadjuvant molecules to improve the transfection rates. One example is the use of cationic lipids (e.g. Lipofectamine®) with a positively charged head group that favours DNA condensation and also facilitates the fusion of the liposome/nucleic acid with the cell membrane prior to the endocytosis.[89]

The first experiments on the use of magnetic fields to enhance nucleic material delivery were first reported by C. Mah et al.,[90] and soon after the term 'magnetofection' was coined by C. Plank's group.[91] Since then, this concept of MNP-mediated transfection has been customized and improved regarding the dose–response ratios and transfection rates. Today, there are many commercially available kits that provide user-ready MNPs and reagents for routine laboratory applications such as CombiMag™ (Ozbiosciences SAS, France) or Magnetofection™ (Chemicell GmbH, Germany).

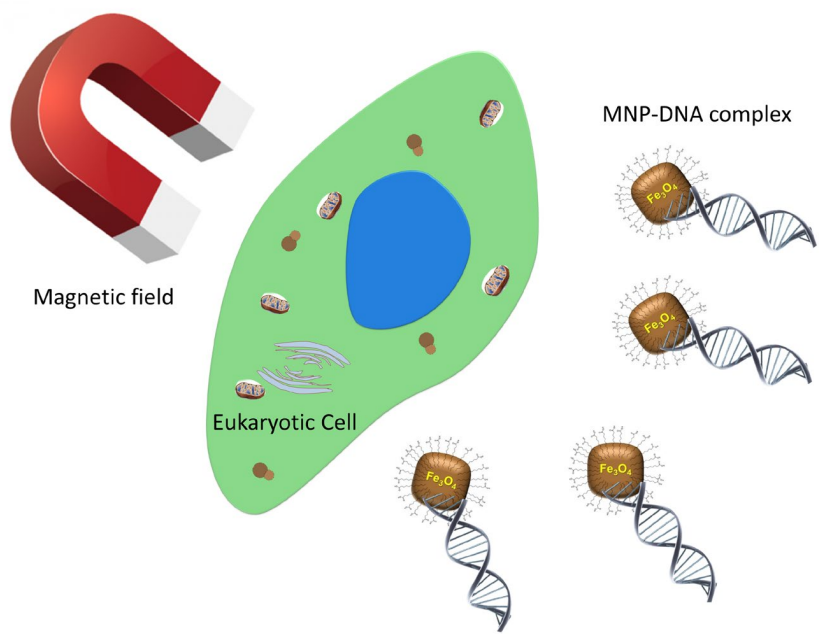

**Figure 22.4:** Schematic illustration of the magnetofection principle. The nucleic acid and the magnetic NP form the magnetic nanovector complex that is pulled towards the cell by a noncontact magnetic force from an external magnetic field gradient. The forces increase the rate of contact events between the vectors and the cell membrane, thus improving the uptake dynamics of the cell.

The basic mechanism is depicted in Figure 22.4: through the use of an external magnetic field gradient, the forces acting on the magnetic vectors increase the contact time of the genetic material and the cell membrane, increasing the uptake dynamics of the cell membrane and thus the efficiency of nucleic acid delivery. Some simple physical models have been proposed for this interaction, based on a drift-diffusion equation through the cell membrane,[92] but a complete model accounting for the different physiological pathways is still lacking. However, there is an emerging consensus that for these applications the surface chemical composition of the MNPs is a key factor determining the final efficiency, irrespective of the details of the magnetic structure of the magnetic cores.

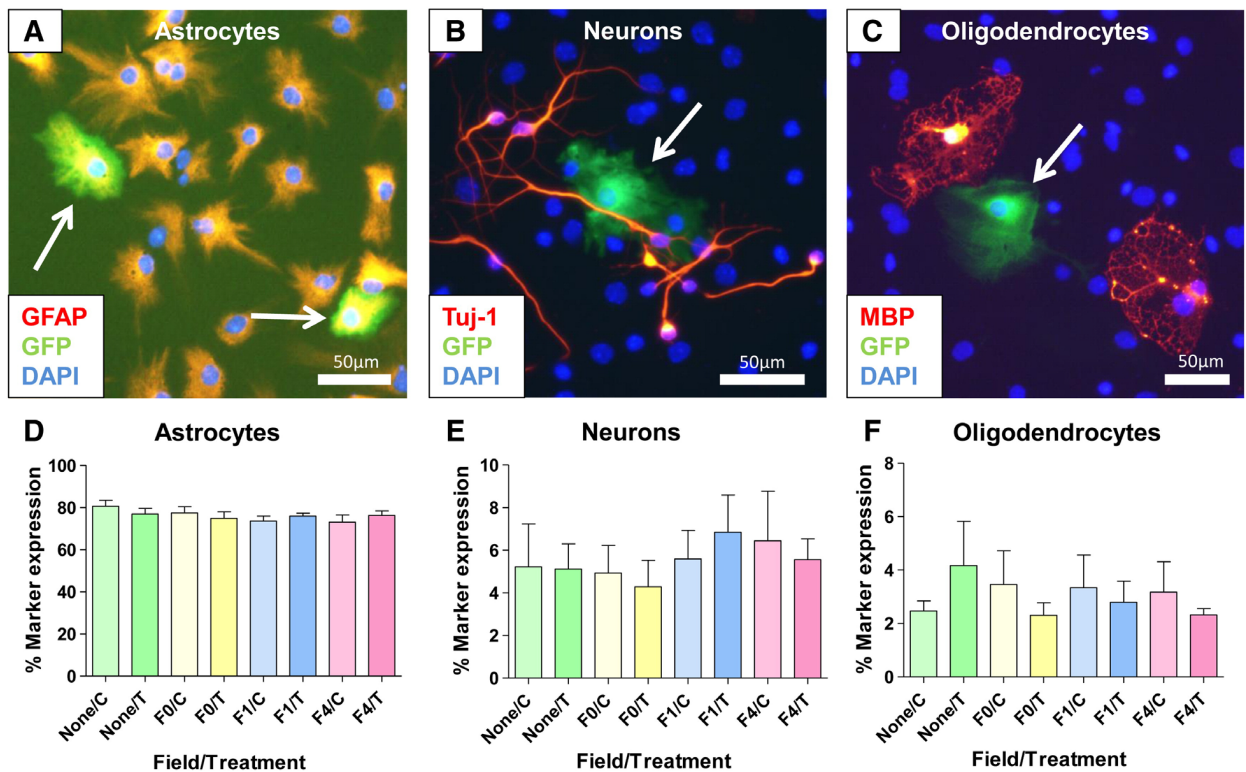

Recently, a method to increase transfection in neural stem cells (NSCs) using MNPs and very low frequency (4 Hz) magnetic fields demonstrated that transfection efficacy could be improved significantly, while keeping the differentiation capabilities unaffected.[93] As shown from the differentiation profiles in Figure 22.5, magnetofected NSCs show positive for all transfected markers. Moreover, the authors reported that magnetofected NSCs displayed disrupted cell membranes as compared to control cells. Although the physical mechanisms involved are not completely understood, experimental data suggest that the higher efficiency under magnetic fields is due to both an increase of the MNP-cell interaction time and a frequency-driven stimulus of the endocytic activity of the cell membrane, as depicted in Figure 22.6.[93]

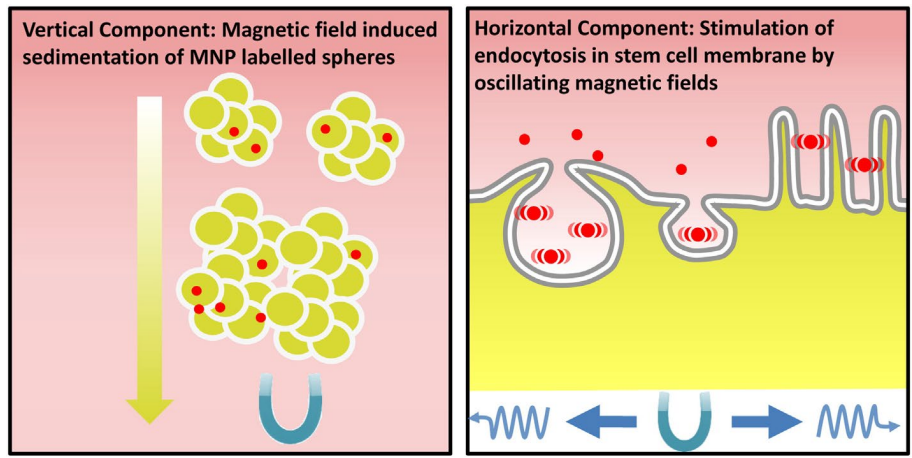

**Figure 22.6:** Proposed mechanism of transfection of neurospheres. Schematic diagram illustrating a hypothetical model to explain the mechanism of oscillating field enhancement of transfection in neurospheres. (Reprinted from *Nanomedicine: Nanotechnology, Biology and Medicine*, 9, C. F.,Adams, M. R., Pickard, and D. M., Chari, 737, Copyright 2013, with permission from Elsevier.)

### 22.4.4 Magnetotransduction

Similarly to the use of magnetic forces, the use of viral vectors has provided a fruitful solution to increase the low efficiency of nonviral gene vectors, a technique known as transduction. The term was coined more than 50 years ago by Zinder et al.,[94] in their genetic studies on *Salmonella typhimurium* and has been improved notably along the last decade.[95] Viral vectors have been intensively used as a tool for fighting NDs, through the delivery of neurotrophic factors that prevent degeneration and enhance recovery of target neurons. The potential of this technique for clinical uses is apparent, especially in the field of NDs. For example, two powerful neuroprotective molecules for the treatment of neurodegenerative pathologies affecting both motor and cognitive functions are GDNF and insulin-like growth factor I (IGF-I).[96] In spite of some promising results, the efficiency of viral (and nonviral) vectors for therapeutic gene delivery into the brain still remains one of the limiting factors to be overcome before clinical trials can be safely implemented. Additionally, most protocols currently in use for nucleic acid delivery[97] require some improvement of either the efficiency or specificity of nucleic acid delivery.[98]

Based on the concepts of magnetofection, i.e. the use of magnetic forces on MNPs to improve transfection efficiency, therapeutic approaches against NDs have begun to use magnetically labelled viral units to deliver genetic material. The construction of magnetic viral vectors (usually adenovirus or lentivirus) for a magnetic field-assisted viral transduction has been reported for some years now. This technique is known as *magnetotransduction*, and is often related (but not restricted) to strategies for delivering neuroprotective molecules to target cells as a therapy against ND diseases. One of the main goals of this approach is related to the enhancement in the levels of neurotrophic factors delivered, since it is accepted that an increase in the delivered concentration of these factors can prevent neural degeneration and enhance recovery of remaining neuron neuroprotective molecules at the target site.[99]

In magnetotransduction, the MNPs also work as the 'pulling' agents when conjugated with viral vectors to construct a magnetic-viral vector of higher efficacy than virus or MNPs alone. Some configurations using $Fe_3O_4$-based MNPs and recombinant adenoviral vector harbouring reporter genes have been already used to magneto-transduce glial and neuronal brain cells (ependymal, hypothalamic and substantia nigra) with high efficiency.[100] Some proof-of-principle experiments with MNP-AAV (adeno-associated viral) vectors showed partial success,[101] but the need for further optimization of vector formulation remains, especially if neuroprotective and neurotrophic factors (e.g. IGF-1, GDNF) are to be used for clinical applications to ND diseases. If successful, this approach could represent a major improvement towards new therapies for NDs.

### 22.4.5 Scavenging Strategies

When the CNS is affected, nerve injury results in a disruption of the blood–spinal cord barrier. Moreover, the damage induced in surrounding blood vessels stimulates a proliferation of Schwann cells, leucocytes, monocytes and macrophages around the nerve lesion that provokes the loss of nervous tissue. At the cellular level, axons show deteriorated myelin layers, and the resulting growth-inhibitory myelin debris is only partially removed by macrophages. Therefore, a containment/scavenging protocol is desired before actual nerve regeneration. Gathering those cells that are activated in response to pathological situations can be achieved by the use of remote magnetic forces on the injured area. The incorporation of MNPs by scavenger cells has been already observed in organotypic culture,[62] and therefore it can be expected that similar targeting can be achieved *in vivo*.[102] Indeed, the *in vitro* preloading of macrophages and the subsequent infiltration *in vivo* for magnetic resonance imaging of injured nerve has been successfully tested some years ago.[103] It is, therefore, a matter of time before similar magnetic labeling of scavenger cells can be used for improved magnetically driven nerve repair. MNPs possess themselves scavenging properties. In particular, their capacity to scavenge free radicals has been used to attenuate oxidative damage induced by $H_2O_2$ in SCI rats when localized by an external magnetic field.[102] Additionally, their functionalization with biomolecules can confer new scavenger capabilities, as recently demonstrated by MNP functionalization with O-methyl-β-cyclodextrin to reduce the extracellular level of L-glutamate in brain nerve terminals.[104]

### 22.4.6 Cell Therapies

Cellular therapies exploit the regenerative potential of cells for nerve repair.[105] They are considered promising, especially for the repair of CNS injuries and long gaps in the PNS. Several cell types such as stem cells, Schwann cells, OECs have been utilized as transplantable cells in nerve regeneration, demonstrating improved regenerative outcomes[106] [107] but, similarly to any cell-based strategy, this approach suffers from drawbacks, which limit the translation from experimental to clinical stages. A great help for implementing safe and effective cell transplantation could be the development of strategies for cell homing and cell tracking, allowing for monitoring of the fate of the transplanted cells and to retain them in the injury/pathology site, maximizing the therapeutic effects while avoiding dangerous migrations to ectopic sites. Several lines of evidence suggest that MNPs hold a great potential to overcome these limitations. Recently, a clinical study[a] has demonstrated in healthy volunteers that MNP can be used for *in vivo* tracking of magnetically labelled human mononuclear cells using MRI scanning. Following intravenous administration, the distribution of iron-labelled cells was monitored as well as their ability to migrate to a site of inflammation. Cell labeling with MNPs can be thus easily imaged via MRI and this approach offers the distinct advantage to correlate the study outcome to the cell localization at the site, or biodistribution in the organism. MNPs have been used to label oligodendrocyte precursor cells, which showed high promise as a transplant population to remyelinate nerve fibres and promote regeneration in the CNS. Indeed, clinical trials using these types of cells have been initiated in some areas.[108] The migration of MNP labelled OPCs was followed via MRI, when injected into the spinal cord of myelin-deficient rats[109] or after transplantation into adult rat brain.[110] Magnetic manipulation is also an advantageous method for guiding cells remotely. Neural progenitor cells[111] or olfactory ensheathing cells[112] have been labelled with MNPs and magnetically localized to promote axon growth in organotypic cocultures. This approach was also used *in vivo* to remotely guide MNP labelled stem cell in the spinal cord of SCI mice, demonstrating enhanced localization and axon regeneration.[130–134]

[a] www.clinicaltrials.gov. Study ref. NCT01169935

## 22.5. Outlook for the Future

There are several nanotherapies already proposed as substitutes for (a) surgical nerve grafting after peripheral nerve injury, (b) pharmacological treatment after drug abuses and (c) neuroprotective drug delivery.[113-115] However, there are no reports to date that can show conclusive clinical improvements over the established surgical procedures. The near future will probably see new nanotherapies as coadjuvant protocols. MNPs have already opened new paths for noninvasive therapies based on the exploitation of the remotely driven mechanical forces on MNP-loaded neurons. The ability of these approaches to promote migration and axonal elongation/growth have already passed the first proof-of-concept challenges, but many fundamental questions are yet unresolved. It is also clear that a 'second generation' of enhanced MNPs is required offering minimum toxicity and better reproducibility. A major issue still not addressed, which will determine the final efficacy of these magnetic vectors, is the creation of a flexible surface for functionalization with neurotrophic/neuroprotective factors. If this flexible platform is developed, it will open boundless possibilities for novel molecular therapies, as well as the basis (together with multipotent stromal cells) for more effective cell therapies.

**Acknowledgements**

This work was partially supported by the Spanish Ministerio de Economia y Competitividad (MINECO) through project MAT2016-78201-P; the Aragon Regional Government (DGA, Project No. E26) and the Wings for Life Spinal Cord Research Foundation (WFL, Project No. 163).

**Dr. Gerardo F. Goya** (*Email: goya@unizar.es) is associate professor at the University of Zaragoza, Spain. He has been associate professor at the University of Sao Paulo (Brazil) and he is currently researcher at the Institute of Nanoscience of Aragón (INA), University of Zaragoza. Prof. Goya's pioneering team (http://www.unizar.es/gfgoya) on magnetic hyperthermia in Spain established that induced cell death with magnetic hyperthermia without temperature rise is possible. His team has developed engineered MNPs for neural guidance under externally applied magnetic fields. He has over 130 publications on nanomagnetism and bioapplications and holds two patents. Prof. Goya 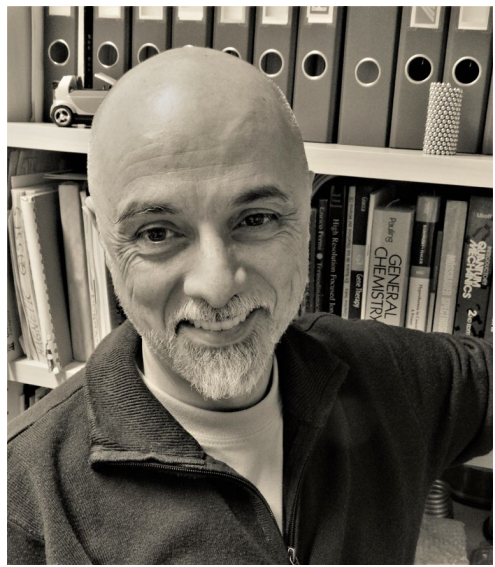 has led the design, development and building of devices for measuring power absorption for magnetic hyperthermia, which established the basis for a spin-off company, nB Nanoscale Biomagnetics, of which he is cofounder and scientific advisor.

**Vittoria Raffa** holds an MSc in Chemical Engineering, PhD in nanotechnology and is associate professor in molecular biology at the University of Pisa. She is the team leader of the Nanomedicine Lab at the University of Pisa (Department of Biology). Prof. Raffa's research interests include nanomedicine and its applications to molecular biology and neuroscience. She was author in the last 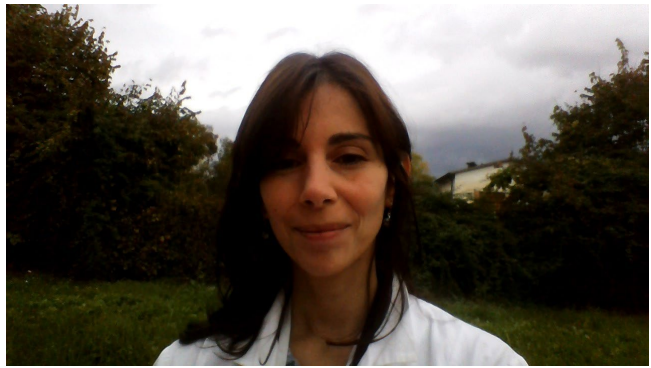 10 years of 60 publications in ISI journals and 5 patents on technologies related to nanomedicine.